\documentclass[aps,pra,twocolumn,showpacs,superscriptaddress]{revtex4}
\usepackage{graphicx,color}
\usepackage{amsmath}
\usepackage{amssymb}

\usepackage{amsfonts}
\usepackage{bm}

\newcommand{\ot}{{\,\otimes\,}}
\newcommand{{\Cd}}{{\mathbb{C}^d}}

\def\oper{{\mathchoice{\rm 1\mskip-4mu l}{\rm 1\mskip-4mu l}%
{\rm 1\mskip-4.5mu l}{\rm 1\mskip-5mu l}}}
\def\<{\langle}
\def\>{\rangle}

\begin{document}
\title{Long-time memory in non-Markovian evolutions }
\author{Dariusz Chru\'sci\'nski}
\affiliation{Institute of Physics, Nicolaus Copernicus University \\
Grudzi\c{a}dzka 5/7, 87--100 Toru\'n, Poland}
\author{Andrzej Kossakowski}
\affiliation{Institute of Physics, Nicolaus Copernicus University \\
Grudzi\c{a}dzka 5/7, 87--100 Toru\'n, Poland}
\affiliation{Dipartimento di Scienze Fisiche and MECENAS, Universit\`a di Napoli ``Federico II", I-80126 Napoli, Italy}
\author{Saverio Pascazio}
\affiliation{Dipartimento di Fisica, Universit\`a di Bari, I-70126 Bari, Italy \\
and Istituto Nazionale di Fisica Nucleare, Sezione di Bari, I-70126 Bari, Italy}

\begin{abstract}
If the dynamics of an open quantum systems is non-Markovian, its
{asymptotic} state strongly depends on the initial conditions, even
if the dynamics possesses an {invariant} state. This is the very
essence of memory effects. In particular, the {asymptotic} state can
remember and partially preserve its initial entanglement.
Interestingly, even if the non-Markovian evolution relaxes to an
equilibrium state, this state needs not be invariant.  Therefore, the
non-invariance of equilibrium becomes a clear sign of non-Markovianity.
\end{abstract}


\maketitle

\section{Introduction}
\label{sec:intro}

Open quantum systems and their dynamical features are attracting increasing attention, nowadays.
Their interest is twofold. On one hand, they are of tantamount importance
in the study of the interaction between a quantum system and its environment,
causing dissipation, decay and decoherence \cite{Weiss,Breuer}.
On the other hand, the robustness of quantum coherence and
entanglement against the detrimental effects of the environment is one of the major scopes in quantum enhanced applications, as both entanglement and quantum coherence are basic resources
in modern quantum technologies, such as quantum communication,
cryptography and computation \cite{QIT}.

The detailed characteristics of the dynamical evolution are far from
being obvious and are often quite surprising. For example, while the
coherence of single qubits in Markovian environments decays
exponentially, the evolution of the entanglement between two qubits
markedly differs and may completely disappear at a \emph{finite}
time (and eventually revive later) \cite{Eberly}, a phenomenon known
as ``entanglement sudden death," that has been recently
experimentally demonstrated \cite{optics} and analyzed from
different perspectives \cite{other}. 

In this paper we will focus on
\emph{non-Markovian} evolutions and will show that they define
a completely new kind of quantum dynamics. In particular this leads to
the modification of the characteristic exponential relaxation law
known from Markovian evolutions. Interestingly, we will show that even
if the non-Markovian evolution relaxes to an equilibrium state this
state needs not be invariant. This can never happen in the Markovian
case. Therefore, the non-invariance of equilibrium becomes a clear
sign of non-Markovianity. It turns out that the {asymptotic} state of
the system depends on the initial conditions, even if the non-Markovian dynamics possesses an {invariant} state. For composed systems this implies that the {asymptotic} states can remember (and partially preserve) its initial entanglement. These conclusions will be illustrated by several examples and pave the way towards a more general comprehension and practical exploitation of non Markovian evolutions.

\section{Preliminary ideas}
\label{sec:prelim}

\subsection{Non-Markovian dynamics}
\label{sec:nonmarkov}

The usual approach to the dynamics of an open quantum system
consists in applying the Markovian approximation, that leads to the
following local master equation
\begin{equation}
\label{M}
   \dot{\rho}_t = L\, \rho_t\ ,
\end{equation}
where $\rho_t$ is the density matrix of the system investigated and
$L$ the time-independent generator of the dynamical semigroup. This
can be formally solved
\begin{equation}
\label{Msol}
   \rho_t  = e^{tL}\rho = \Lambda_t \rho  \qquad (t \geq 0, \rho=\rho_{t=0})
\end{equation}
and it is well known that under certain conditions on $L$ \cite{GKSL}
the dynamics $\Lambda_t$ is completely positive and trace preserving
\cite{Alicki,Breuer}.

Let us study the behavior of quantum coherence under non-Markovian
evolutions. For the sake of simplicity, we shall restrict our
attention to finite level systems. A popular non-Markovian
generalization of (\ref{M}) is the following nonlocal equation
\begin{equation}\label{NM}
   \dot{\rho}_t = \int_0^t L_{t-\tau}\, \rho_\tau \, d\tau\ ,
\end{equation}
in which quantum memory effects are taken into account through the
introduction of the memory kernel $L_\tau$: this simply means that
the rate of change of the state $\rho_t$ at time $t$ depends on its
history (starting at $t=0$). The Markovian master equation (\ref{M})
is reobtained when $L_\tau = 2\delta(\tau)L$. The time dependent
kernel $L_\tau$ is usually referred to as the generator of the
non-Markovian master equation. Equation (\ref{NM})
applies to a variety of situations, e.g.\ when the particle is
born in the medium in which it propagates (neutrinos in a stellar
medium \cite{Lisi} or pairs of neutral kaons in the gravitation
field of a laboratory \cite{benatti}).

One of the fundamental problems in the theory of non-Markovian
master equations is to find those conditions on $L_\tau$ that ensure
that the time evolution resulting from (\ref{NM})
\begin{equation}\label{}
   \rho \ \longrightarrow\ \rho_t = \Lambda_t \rho
\end{equation}
is completely positive (CP) and trace preserving
\cite{Wilkie,Budini,B-2004,Lidar,Maniscalco1,Maniscalco2,KR,B}. {Let
us observe  that this problem may be reformulated as follows
\cite{KR-last}:  any completely positive solution $\Lambda_t$ of Eq.
(\ref{NM}) may be represented by
\begin{equation}\label{Phi}
   \Lambda_t = \oper + \int_0^t \Phi_\tau \, d\tau\ ,
   \end{equation}
where the maps $\Phi_\tau$ satisfy ${\rm Tr}\,\Phi_\tau \rho=0$ for
all $\rho$. This condition guaranties that $\Lambda_t$ is trace
preserving.   It is easy to show that the Laplace transform of the
generator $L_\tau$ of the non-Markovian master equation (\ref{NM}) is
related to the Laplace transform of $\Phi_\tau$ as follows
\begin{equation}\label{Laplace-L}
   \widetilde{L}_s = \frac{s \widetilde{\Phi}_s}{ \oper +
   \widetilde{\Phi}_s}\ .
\end{equation}
Now, in order to explicitly write down $L_\tau$ one has to invert the
Laplace transform $\widetilde{L}_s$. Note, however, that this might
be very hard, due to the fact that $\widetilde{L}_s$ is a highly
nontrivial function of $s$ (possessing in general not only poles but
also cuts in the complex $s$-plane). It is therefore clear that even
if one knows the solution $\rho_t = \Lambda_t \rho $, it is in
general very difficult (if not practically impossible) to write down
the corresponding non-Markovian equation (\ref{NM}). On the other
hand, the knowledge of the (trace preserving and CP) solution
$\Lambda_t$ enables one to no longer care about the underlying
equation! Let us look at an interesting example.

\subsection{An example}
\label{sec:example}

The previous comments are best understood by looking at an example.
Consider the pure decoherence model,
\begin{equation}
\label{Htot}
H=H_R + H_S + H_{SR},
\end{equation}
where
$H_R$ is the reservoir Hamiltonian,
\begin{equation}
\label{HS}
H_S = \sum_n \epsilon_n P_n \; \quad (P_n=|n\>\< n|)
\end{equation}
the system Hamiltonian and
\begin{equation}\label{}
   H_{SR} = \sum_n P_n \ot B_n
\end{equation}
the interaction part, respectively, $B_n=B_n^\dagger$ being reservoirs operators.
The initial product state $\rho \ot \omega_R$ evolves according to the
unitary evolution $e^{-i H t} (\rho \ot \omega_R) e^{i H t}$ and by
partial tracing with respect to the reservoir degrees of freedom one
finds for the evolved system density matrix
\begin{eqnarray}
   \rho_t &=&
   {\rm Tr}_R [e^{-i H t} (\rho \ot \omega_R) e^{i H t}] =
   \sum_{n,m} c_{mn}(t)P_m\rho P_n \ , \nonumber \\
   \label{rho0}
   \end{eqnarray}
where
\begin{eqnarray}
   c_{mn}(t) = {\rm Tr}( e^{-iZ_m t}\omega_R e^{iZ_n t}),
\end{eqnarray}
and the reservoir operators $Z_n$ are defined by
\begin{eqnarray}
& & Z_n= \epsilon_n \mathbb{I}_R + H_R + B_n .
\end{eqnarray}
Note that the matrix $c_{mn}(t)$ is semi-positive definite and hence
Eq.\ (\ref{rho0}) defines the Kraus-Stinespring representation
\cite{Kraus} of the completely positive map $\Lambda_t$
\begin{equation}\label{}
   \Lambda_t \rho  = \sum_{n,m} c_{mn}(t) P_m \rho P_n\ .
\end{equation}
The prescription (\ref{Phi}) yields
\begin{eqnarray}\label{}
 & &  \rho_t = \rho + \int_0^t \sigma_\tau\,d\tau\ , \\
& & \sigma_\tau = \Phi_\tau \rho  = \dot{\rho}_\tau = \sum_{n,m} \dot{c}_{mn}(\tau)
P_m \rho P_n \
\end{eqnarray}
and one very easily shows that ${\rm Tr}\,\sigma_\tau = 0$. The
solution of the pure decoherence model can therefore be found
without explicitly writing down the underlying master equation. What
is (and needs to be) known is that $\rho_t$ satisfies the
non-Markovian master equation (\ref{NM}), but the construction of
the corresponding memory kernel $L_t$ is too formidable a task.
{Indeed, let us observe that due to the following spectral property
of $\Lambda_t$
\begin{equation}\label{}
   \Lambda_t |m\>\<n| = c_{mn}(t)  |m\>\<n| \ ,
\end{equation}
one obtains the following formula for the corresponding generator
\begin{equation}\label{}
L_t \rho  = \sum_{n,m} \kappa_{mn}(t) P_m \rho P_n\ ,
\end{equation}
where the functions $\kappa_{mn}(t)$ are defined in terms of their
Laplace transform as follows
\begin{equation}\label{}
   \widetilde{\kappa}_{mn}(s) = \frac{s\widetilde{c}_{mn}(s) -
   1}{\widetilde{c}_{mn}(s)} \ .
\end{equation}
Note, that $c_{mm}(t)=1$, and hence $\kappa_{mm}(t)=0$. This
condition guaranties that $L_t \mathbb{I}=0$. However, the
calculation of the off--diagonal elements $\kappa_{mn}(t)$ is in
general not feasible.

Many similar examples are known in the physical literature, e.g.\ in
connection with the quantum Zeno effect. See \cite{zenorev} for a
review on non Markovian decay and \cite{Raizen} for its experimental
observation. In the following we shall therefore work directly with
$\Lambda_t$ and Eqs.\ (\ref{Phi})-(\ref{Laplace-L}), without
detailing the features of the appropriate memory kernel $L_t$.

\section{Asymptotic vs equilibrium states}
\label{sec:asympequil}

Let us now point out the crucial difference between Markovian and
non-Markovian evolutions. Recall that a state $\omega$ is an
{equilibrium} state for the (Markovian or non-Markovian) evolution
$\Lambda_t$ if
\begin{equation}
\label{eqstate} \lim_{t \to \infty }\Lambda_t \rho  = \omega \quad
\forall \rho.
\end{equation}
One says that the evolution relaxes to $\omega$ and
we shall assume for simplicity that $\omega$ is unique for the given $\Lambda_t$.
On the other hand
a state $\rho_0$ is an {invariant} state for $\Lambda_t$ if
\begin{equation}\label{omega-inv}
   \Lambda_t \rho_0 =\rho_0\ \ \ \forall t\geq 0\ .
\end{equation}
Note that if $\Lambda_t$ defines a semigroup, i.e. $\Lambda_t =
e^{tL}$, then $\rho_0$ is {invariant} if $L \rho_0=0$. Clearly, for
Markovian evolution the {equilibrium} state $\omega$ is always
{invariant}. This is a straightforward consequence of the semigroup
property $\Lambda_{s+t}(\omega) = \Lambda_t(\Lambda_s(\omega))$ in
the limit $s \rightarrow \infty$. However, this property is no
longer true in the non-Markovian case, where the semigroup property
cannot be used. Therefore, one may have non-Markovian evolutions
relaxing to an {asymptotic} {equilibrium} state which is not
{invariant}. In the following, we shall analyze a few situations in
order to explore the relaxing properties of non-Markovian
evolutions.

\subsection{A case study: convex combination of Markovian semigroups}
\label{sec:casestudy}

Let $L_1,\ldots,L_n$ be a set of generators of
Markovian equations of the type (\ref{M}) and let $(p_1,\ldots,p_n)$
be a probability distribution $(\sum p_k=1)$. Then
\begin{equation}\label{pL}
   \Lambda_t = \sum_{k=1}^n p_k \, e^{tL_k}\ ,
\end{equation}
is by construction completely positive and satisfies (\ref{Phi})
with
\begin{equation}\label{}
   \Phi_t =  \frac{d \Lambda_t}{d t}= \sum_{k=1}^n p_k \, L_k e^{tL_k} \ .
\end{equation}
Actually, it is not difficult to conceive an evolution that is a
convex combination of Markovian semigroups.
Consider a system
$S$ living in $\mathcal{H}_S$ coupled to a reservoir $R$ living in
$\mathcal{H}_R$. (Actually, one may consider an arbitrary number $N$
of reservoirs. In this case $\mathcal{H}_R =
\mathcal{H}_1 \ot \ldots \ot \mathcal{H}_N$.) Now, couple the
composed $S$-$R$ system  to an $n$-level ancilla living in
$\mathbb{C}^n$ and assume that the Hamiltonian has the following
form
\begin{equation}\label{H-total}
   H = \sum_{k=1}^n H_k \ot P_k ,
\end{equation}
where $P_k = |k\>\<k|$ ($|k\>$ is an orthonormal basis in
the ancilla Hilbert space $\mathbb{C}^n$) and $H_k=H_k^\dagger$ are
$S$-$R$ operators. The unitary evolution generated by
(\ref{H-total}) reads
\begin{equation}
e^{-itH}=\sum_{k=1}^n e^{-itH_k} \ot P_k,
\end{equation}
hence if the initial product state is $\rho \ot \omega_{R}\ot
\sigma$, $\sigma$ being a state of the ancilla, the reduced dynamics  yields the following evolution  for
the system density operator
\begin{equation}\label{rho}
  \rho_t = \sum_{k=1}^n  p_k {\rm Tr}_{R} [e^{-i H_k t} (\rho \ot \omega_{R}) e^{i H_k t}] ,
\end{equation}
where $p_{k} = \langle k|\sigma|k\rangle$. Standard weak coupling
arguments lead to (\ref{pL}).

A convex combination (\ref{pL}) of Markovian semigroups is no longer
a semigroup and satisfies the non-Markovian master equation
(\ref{NM}). However, it can be very complicated to find the
corresponding memory kernel. Observe that if for each $k$ the
corresponding Markovian evolution $\Lambda^{(k)}_t = e^{tL_k}$
possesses a unique equilibrium (and hence {invariant}) state $\omega_k$,
then $\Lambda_t$ defined by (\ref{pL}) relaxes to the {equilibrium}
state $\omega = \sum_{k=1}^n p_k \omega_k$. Note that $\omega_k$
need not be {invariant} for $\Lambda^{(l)}_t$ with $\l \neq k$ (it
is {invariant} if $L_l$ and $L_k$ commute). We stress that if each
subgroup of ensemble members has its own Markovian decay process, towards
its own equilibrium, then the global (non-Markovian) dynamics has a well defined
equilibrium (convex combination of Markovian equilibria) and hence
the final  state does not depend on the initial state (by 
definition of equilibrium). However, the {equilibrium} state
$\omega$ needs not be {invariant} for the non-Markovian evolution
governed by (\ref{pL}). That is, in general $\Lambda_t\omega \neq
\omega$, but of course asymptotically $\lim_{t\rightarrow\infty}
\Lambda_t \omega = \omega$.}

The simplest
example of (\ref{pL}) corresponds to $L_1=L$ and $L_2=0$, yielding
the following non-Markovian evolution
\begin{equation}\label{I-Lambda}
   \Lambda_t = (1-p)e^{tL} + p \oper\ ,
\end{equation}
i.e.\, a mixture of a semigroup dynamics $e^{tL}$ and the trivial
one $\oper \rho =\rho$. Equations (\ref{Phi})-(\ref{Laplace-L})
yield
\begin{equation}\label{}
   \widetilde{L}_s = (1-p)L + \frac{p(1-p)\, L^2}{s - (1-p)L}\ ,
\end{equation}
which can be easily inverted
\begin{equation}\label{I-L}
   L_t = 2(1-p)\delta(t)L + p(1-p)L^2 e^{t(1-p)L}\ .
\end{equation}
Note the similarity with the Shabani-Lidar \cite{Lidar} memory
kernel $L_t = Le^{tL}$ of the post-Markovian quantum master
equation.  In general $Le^{tL}$  does not lead to a completely
positive dynamics. On the other hand, the kernel (\ref{I-L})
generates a completely positive dynamics for arbitrary $L$. Formula
(\ref{I-Lambda}) is an exceptional case: in general one cannot
obtain a closed expression for the generator $L_t$. We stress that
the non-Markovian dynamics (\ref{I-Lambda})  displays very peculiar
features. Suppose that $e^{tL}$ possesses an {equilibrium} (and
hence {invariant}) state $\omega$. It is clear that $\omega$ is
still {invariant} for (\ref{I-Lambda}) but it is no longer an
{equilibrium} state. Note, that $L_t\omega=0$ due to the fact that
$L\omega=0$. In conclusion, one has
\begin{equation}\label{As-0}
\lim_{t\rightarrow \infty} \Lambda_t \rho  = (1-p) \omega + p\rho ,
\end{equation}
which shows that $\omega$ cannot be reached {asymptotic}ally (unless
we start with $\omega$ itself). Since, in general, a non-Markovian
evolution is not relaxing, the {asymptotic} state strongly depends
on the initial condition.  This is the very essence of memory
effects---the system remembers its initial state. We stress that
this result is model independent. The only assumption is that $L$
generates a relaxing Markovian semigroup. For example one may take
instead of the trivial generator $L_2=0$ the following one
\begin{equation}
\label{Lprime} L_2' = - \gamma (\oper - {\cal P}), \quad \gamma \geq
0 ,
\end{equation}
where
\begin{equation}
\label{PP} {\cal P} \rho  = \sum_n P_n \rho P_n
\end{equation}
is a projector, with
$P_n=|n\>\<n|$, $|n\>$ being eigenvectors of $\omega$. One has
therefore ${\cal P}\omega=\omega$. Hence, the convex combination
(\ref{pL}) yields  the following formula
\begin{equation}\label{I-Lambda-new}
   \Lambda'_t = (1-p)e^{tL} + p\Big[\mathcal{P} + e^{-\gamma t}( \oper -
   {\cal P})\Big] .
\end{equation}
For $\gamma=0$, $L_2' = L_2$ and one recovers (\ref{As-0}). For
$\gamma > 0$ the {asymptotic} formula (\ref{As-0}) is replaced by
\begin{equation}\label{As-1}
\lim_{t\rightarrow \infty} \Lambda'_t \rho  = (1-p) \omega +
p\,{\cal P} \rho  .
\end{equation}
Again, $\omega$ defines an {invariant} state for $\Lambda'_t$.
However, $\Lambda_t'$ is not relaxing and $\omega$  is not reachable
(unless we start from it). Observe that the mixing parameter $p\in
[0,1]$ in (\ref{I-Lambda}) and (\ref{I-Lambda-new}) measures in a
sense the ``non-Markovianity" of the evolution.

\subsection{Quantum channel}
\label{sec:qchannel}

We now look at a different example. Let
\begin{equation}\label{Lt}
   L_t = \kappa(t) \, ({B} - \oper)\ ,
\end{equation}
where $B$ is a quantum channel (i.e., a trace preserving CP map) \cite{Budini,B,KR-last}.
$L_t$ generates a completely positive trace preserving dynamics $\Lambda_t$ if the Laplace transform $\widetilde{\kappa}(s)$ satisfies
\begin{equation}\label{k-f}
   \widetilde{\kappa}(s) =
   \frac{s\widetilde{f}(s)}{1-\widetilde{f}(s)}\ ,
\end{equation}
where $f(t) \geq 0$ and $
   \int_0^\infty f(\tau)d\tau \leq 1\, .$  Note that the
corresponding Laplace transform of $\Lambda_t$ reads
\begin{equation}
\label{tilde-Lambda}
\widetilde{\Lambda}_s = \frac{1}{s}
\frac{1-\widetilde{f}(s)}{\oper-\widetilde{f}(s)B}
\end{equation}
and in general cannot be inverted.  However, even if we are not able
to find $\Lambda_t$, we can easily study its {asymptotic} behavior.
Indeed, using the well known property of the Laplace transform
\begin{equation}\label{laplprop}
   \lim_{t \rightarrow \infty} \Lambda_t = \lim_{s\rightarrow 0}
   s\widetilde{\Lambda}_s\ ,
\end{equation}
if all poles of $s\widetilde{\Lambda}_s$ are in the left-hand plane,
one obtains from (\ref{tilde-Lambda}) the general {asymptotic} formula
\begin{equation}\label{Asymp}
\Lambda_\infty =
\frac{1-\widetilde{f}(0)}{\oper-\widetilde{f}(0)B}\ .
\end{equation}
To study $\Lambda_\infty$ in more detail consider the spectral
decomposition of $B$:
\begin{equation}\label{}
   B \rho  = \sum_{\alpha=0}^{d^2-1} b_\alpha F_\alpha
{\rm Tr}(G_\alpha^\dagger \rho) \ ,
\end{equation}
where $d$ stands for the dimension of the system Hilbert space, and
$F_\alpha$ and $G_\alpha$ define the bi-orthogonal damping basis of
$B$. Suppose now that $B$ possesses the unique {invariant} state
$\rho_0$. This implies $F_0=\rho_0$, $G_0=\mathbb{I}$ and the
corresponding eigenvalue $b_0=1$. One has therefore
\begin{equation}\label{}
{\Lambda}_\infty \rho   =  \rho_0 + \sum_{\alpha=1}^{d^2-1}
\frac{1-\widetilde{f}(0)}{1-\widetilde{f}(0)b_\alpha} \,F_\alpha
{\rm Tr}(G_\alpha^\dagger \rho) \ .
\end{equation}
Let us observe that if
\begin{equation}\label{}
\widetilde{f}(0)= \int_0^\infty
f(\tau)d\tau =1,
\end{equation}
then $\Lambda_\infty \rho =\rho_0$, that is, the non-Markovian
dynamics $\Lambda_t$ is relaxing to the {asymptotic} {equilibrium}
state $\rho_0$. However, if $\widetilde{f}(0)<1$, then {the dynamics
is no longer relaxing and the} {asymptotic} state $\Lambda_\infty
\rho $ remembers {about the}  initial state $\rho$.

{Consider for example $f(\tau) = \varepsilon\gamma e^{-\gamma
\tau}$, with $\gamma >0$ and $\varepsilon \in (0,1]$. One has in
this case
\begin{equation}\label{}
\widetilde{f}(0)= \varepsilon \leq 1  ,
\end{equation}
and hence the parameter $\varepsilon$ controls the asymptotic
state $\Lambda_\infty\rho$. Let us observe that one can easily invert the
Laplace transform  (\ref{k-f}) to obtain the following expression for the
function $\kappa(t)$:
\begin{equation}\label{}
   \kappa(t) = \varepsilon\gamma\Big[ 2 \delta(\tau) - \gamma(1-\varepsilon)
   e^{-\gamma(1-\varepsilon)\tau}\Big] .
\end{equation}
Observe that for $\varepsilon=1$, one gets
$\kappa(t)=2\gamma\delta(t)$ which corresponds to the Markovian
dynamics. Hence, the parameter $1-\varepsilon$ measures the
deviation from the Markovianity.

This shows that non-Markovian evolutions are much more flexible. One can
control the asymptotic behavior by controlling a single function of time
$f(t)$ (for example by controlling a single parameter $\varepsilon$).
Note that in the Markovian case the evolution generated by
(\ref{Lt}) is given by
\begin{equation}\label{}
   \Lambda_t^{\rm M}\rho = \sum_\alpha e^{\gamma b_\alpha t} F_\alpha {\rm Tr}(G_\alpha^\dagger \rho) ,
\end{equation}
and hence it displays the characteristic exponential behavior
$\exp(\gamma b_\alpha t)$. We stress that the Markovian evolution is
relaxing to the unique invariant state $\rho_0$, i.e.\ $\rho_0$ plays
the role of equilibrium state for $\Lambda_t^{\rm M}$. In the
non-Markovian case the evolution is relaxing only if
$\widetilde{f}(0)=1$. Note, however, that even if the evolution is
relaxing, relaxation needs not be exponential.

\subsection{Entanglement}
\label{sec:entang}

It is clear that if $e^{tL}$ describes the {relaxing} evolution of a
composed system and its {equilibrium} state $\omega$ is separable,
then all initially entangled states {asymptotic}ally become
disentangled. This is no longer true for non-Markovian evolutions,
such as (\ref{I-Lambda}) and (\ref{I-Lambda-new}). Whether the
{asymptotic} state is separable or not may depend on the initial
state as well. If one starts at time $t=0$ with an entangled state
$\rho$, the {asymptotic} state (\ref{As-0}) or (\ref{As-1}) might be
entangled even if $\omega$ is separable. Moreover the system may consists of an arbitrary
number of parties. For example, in the simplest case of a 2-qubit
system possessing an invariant (but not equilibrium) state
$\omega$ which is maximally mixed, i.e.\ $\omega = \mathbb{I}/4$,
Eq.\ (\ref{As-0}) defines a mixed {asymptotic} state
$(1-p)\mathbb{I}/4 + p\rho$. Hence, starting with a maximally
entangled  state $|\psi\>$ the dynamics (\ref{I-Lambda})
{asymptotic}ally approaches a Werner-like state
\begin{equation}\label{}
   \frac{1-p}{4}\,\mathbb{I} + p |\psi\>\<\psi|  ,
\end{equation}
which is entangled if $p> 1/3$ \cite{Werner}. Hence, the
``non-Markovianity parameter" $p$ controls the entanglement of the
{asymptotic} state.

Similarly, using the spectral resolution $\mathbb{I} = \sum_\alpha
|\psi_\alpha\>\<\psi_\alpha|$, with $|\psi_\alpha\>$ being the four
Bell states, one finds that starting with an initial state $\rho$
the non-Markovian dynamics (\ref{I-Lambda-new}) with an {invariant}
state $\omega = \mathbb{I}/4$ {asymptotic}ally approaches the
Bell-diagonal state
\begin{equation}\label{BD}
   \sum_\alpha p_\alpha |\psi_\alpha\>\<\psi_\alpha| ,
\end{equation}
with $p_\alpha = (1 - p)/4 + p\, \<\psi_\alpha|\rho|\psi_\alpha\>$
depending upon the initial state $\rho$. It is well known that
(\ref{BD}) is entangled if exactly one  $p_\alpha > 1/2$. Again, $p$
controls the separability properties of the {asymptotic} state
(\ref{BD}).

Finally, consider the non-Markovian dynamics generated by the generator
(\ref{Lt}), where $B$ is a quantum channel $B :
\mathcal{B}(\mathcal{H}_1 \ot \mathcal{H}_2) \longrightarrow
\mathcal{B}(\mathcal{H}_1 \ot \mathcal{H}_2)$. The simplest example
of $B$ is a projection defined by
\begin{equation}\label{}
 B\rho = \sum_{m,n} P_{mn} \rho P_{mn} ,
\end{equation}
where $P_{mn} = |m\ot n\>\<m \ot n| = P_m \ot P_n$ are projectors
onto the product vectors of the orthonormal basis in $\mathcal{H}_1
\ot \mathcal{H}_2$. Hence, if $\rho$ is a density operator of the
bi-partite system living in $\mathcal{H}_1 \ot \mathcal{H}_2$, then
representing $\rho$ in the block form
\begin{equation}\label{}
   \rho = \sum_{m,n} |m\>\<n| \ot \widehat{\rho}_{mn} ,
\end{equation}
where $\widehat{\rho}_{mn}$ are operators in
$\mathcal{B}(\mathcal{H}_2)$, one finds for the action of the
projection $B$
\begin{equation}\label{B-rho}
   B\rho = \sum_{m,n} (\widehat{\rho}_{nn})_{mm} P_m \ot P_n .
\end{equation}
It is  easy to find the solution of the non-Markovian master
equation
\begin{equation}\label{}
   \Lambda_t = \left(1 - \int_0^t f(\tau) d\tau\right) \oper + \int_0^t
   f(\tau)d\tau B  ,
\end{equation}
where $f(\tau)$ is defined via formula (\ref{k-f}). The density matrix has the following behavior: the diagonal blocks read
\begin{eqnarray}\label{}
\widehat{\rho}_{mm}(t) &=& \left(1 - \int_0^t f(\tau) d\tau\right)\,
   \widehat{\rho}_{mm} \nonumber  \\ & & + \int_0^t f(\tau)d\tau\, \sum_k
   (\widehat{\rho}_{kk} )_{mn} P_k ,
\end{eqnarray}
and the off-diagonal blocks
\begin{equation}\label{}
   \widehat{\rho}_{mn}(t) =\left(1 - \int_0^t f(\tau) d\tau\right)\,
   \widehat{\rho}_{mn} ,
\end{equation}
for $m\neq n$. This shows that during the evolution the off-diagonal
blocks are scaled by the factor $1- \int_0^t f(\tau)d\tau$ and
eventually disappear if $\int_0^\infty f(\tau)d\tau=1$. The
asymptotic state of the bi-partite system reads 
\begin{equation}\label{}
   \Lambda_\infty \rho = (1- \widetilde{f}(0)) \rho +
   \widetilde{f}(0) B\rho .
\end{equation}
The asymptotic entanglement is controlled by $\widetilde{f}(0)$. It
is therefore clear that if $\widetilde{f}(0)=1$, then
\begin{equation}\label{}
\Lambda_\infty \rho =  B\rho ,
\end{equation}
which is separable being block-diagonal (the off-diagonal blocks
disappear). Actually, due to formula (\ref{B-rho})  the
asymptotic state $B\rho$ is not only block diagonal but even
diagonal in the $|m \ot n\>$ basis. It is, therefore, clear that in
this case the state becomes separable in finite time and hence one
encounters the sudden death of entanglement \cite{Eberly}. This happens in particular in the Markovian case (for a Markovian evolution one has
$1- \int_0^t f(\tau) d\tau = e^{-\gamma t}$).  However, taking
$f(\tau) = \varepsilon \gamma e^{-\gamma \tau}$ one has
$\widetilde{f}(0) = \varepsilon$, and hence
\begin{equation}\label{}
   \Lambda_\infty \rho = (1- \varepsilon) \rho +
   \varepsilon B\rho ,
\end{equation}
which shows that $\varepsilon$ can control the asymptotic
entanglement of $\rho_\infty$. Starting from an entangled $\rho$ one
may preserve entanglement forever by taking a large enough deviation $1-\varepsilon$ from Markovianity.

\section{Conclusions}
\label{sec:concl}

In conclusion, we have shown that non-Markovian dynamics represent
a completely new kind of quantum evolution. They are much more flexible
than the Markovian ones and can lead to a completely
novel behavior of the quantum system. In general, they provoke a modification of the characteristic exponential relaxation law known
from Markovian evolutions. As a consequence, non-Markovianity entails new
features of decoherence and relaxation to equilibrium.
Interestingly, even if the evolution relaxes to an
equilibrium state, this state need not be invariant. This can never
happen in the Markovian case. Therefore, the non-invariance of 
equilibrium becomes a clear sign of non-Markovianity.

We have shown the {asymptotic} state of the system depends on the
initial conditions, even if the non-Markovian dynamics possesses an
{invariant} state. For composed systems this implies that the
{asymptotic} states can remember (and partially preserve) its
initial entanglement. Hence some residual entanglement can remain
even in the remote future. Therefore, non-Markovian evolutions may
avoid the phenomenon of sudden death of entanglement and can preserve
entanglement forever. Our examples show that the  asymptotic
entanglement can be controlled by some characteristic parameters of
the system in question (we  called them non-Markovianity parameters).
These model-independent conclusions have been illustrated by several
examples and seem to pave the way towards a more general
comprehension and practical exploitation of non Markovian
evolutions.

This work was partially supported by the Polish Ministry of Science
and Higher Education Grant No 3004/B/H03/2007/33  and by the EU
through the Integrated Project EuroSQIP. SP would like to thank the
Institute of Physics of the Nicolaus Copernicus University for their
warm hospitality.

\end{document}